\documentclass{epl}
\usepackage{psfig}

\institute{
\inst{1} Institut de Physique de la Mati\`{e}re Complexe,
Facult\'{e} des Sciences de Base, Ecole Polytechnique F\'{e}d\'{e}rale
de Lausanne, CH-1015 Lausanne, Switzerland \\
\inst{2} Laboratoire de
Physique Th{\'e}orique Unit{\'e} Mixte de Recherche UMR 8627,
B{\^a}timent 210, Universit{\'e} de Paris-Sud, 91405 Orsay Cedex,
France \\
\inst{3} Universit\`a degli Studi di Roma ``La Sapienza'',
Dipartimento di Fisica, P.le A. Moro 5, 00185 Rome, Italy and INFM,
Center for Statistical Mechanics and Complexity, Rome, Italy}
\begin{document}

\title{Extreme events driven glassy behaviour in granular media}
\author{G. D'Anna\inst{1}, P. Mayor\inst{1}, G. Gremaud\inst{1}, A. Barrat\inst{2}
and V. Loreto\inst{3}}
\maketitle

\begin{abstract}
Motivated by recent experiments on the approach to jamming of a weakly
forced granular medium using an immersed torsion oscillator [Nature \textbf{
413}, 407 (2001)], we propose a simple model which relates the microscopic
dynamics to macroscopic rearrangements and accounts for the following
experimental facts: (1) the control parameter is the spatial amplitude of
the perturbation and not its reduced peak acceleration; (2) a
Vogel-Fulcher-Tammann-like form for the relaxation time. The model draws a
parallel between macroscopic rearrangements in the system and extreme events
whose probability of occurrence (and thus the typical relaxation time) is
estimated using extreme-value statistics. The range of validity of this
description in terms of the control parameter is discussed as well as the
existence of other regimes.
\end{abstract}

\section{Introduction}

Granular matter physics~\cite{RMP,exp} is a very interesting laboratory for
addressing open problems of non-equilibrium statistical mechanics, such as
the nature of slow glassy dynamics~\cite{exp-chicago} and jamming~\cite
{jamming}, the mechanisms of pattern formation, the physics of avalanche
phenomena. In this context, recent experimental results have shown~\cite
{danna-nature,danna-prl} the intriguing analogy between the way a perturbed
granular medium progresses towards complete rest by decreasing the amplitude
of external taps, and the vitrification of glass-forming materials.

This slow glass-like granular behavior raises one of the main questions in
the framework of granular matter, namely the link between the macroscopic
response to an external perturbation and the microscopic (or mesoscopic)
processes occurring at the scale of the single particle. Recently, it has
been conjectured ~\cite{danna-pre} that the occurrence probability of 
\textit{rare events} might explain the origin of the slow dynamic behavior
of granular matter. Such rare events arise when the largest
externally-induced vibration in the granular medium overcomes some suitable
threshold, triggering in this way a macroscopic ``fracture`` of the granular
solid, that is a grain rearrangement. In this paper we further develop this
idea and provide a model close to the intuitive picture given by experiments
and which may be of interest for similar systems such as glasses or
disordered systems~\cite{rammal,vinokur,bouchaud}.

\section{Experimental results}

\begin{figure}[tbh]
\centerline{
\psfig{file=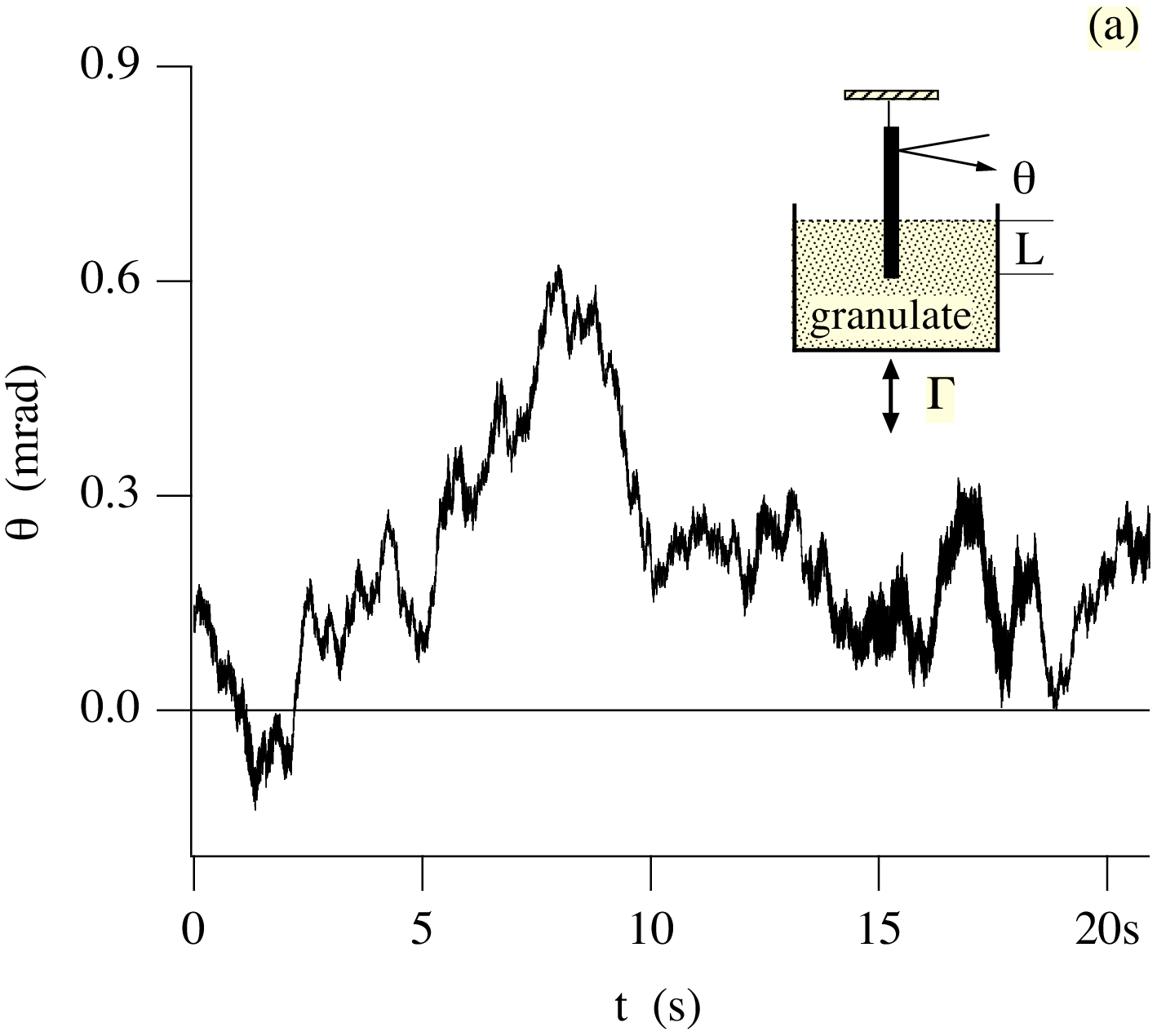,width=6.0cm,angle=0}
\psfig{file=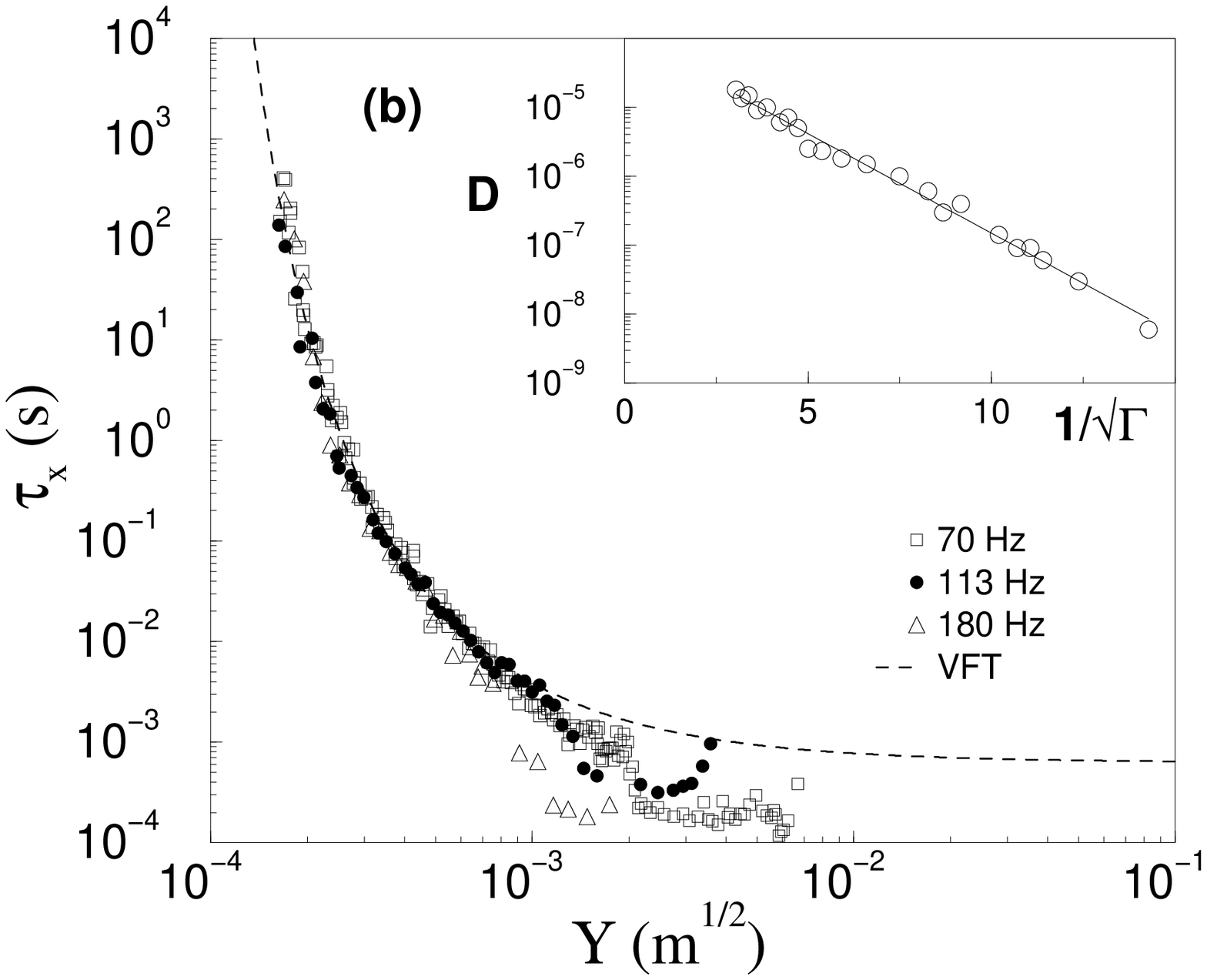,width=6.0cm,angle=0}}
\vspace{0.2cm}
\caption{(a): Typical time series $\theta (t)$ obtained at $\Gamma =0.056$
and $f_{s}=113$ Hz.. Inset: Sketch of the immersed torsion oscillator. (b):
Relaxation time $\tau _{x}$ vs. $\Upsilon $ for various forcing frequencies 
$f_{s}$. Dashed line: VFT fit $\tau _{x}=A\exp [B/(\Upsilon -\Upsilon
_{0})^{p}]$ with $A=6.5\times 10^{-4}$, $B=0.0072$, $p=0.95$, $\Upsilon
_{0}=1.37\times 10^{-4}m^{1/2}$. Inset: Diffusion coefficient (arbitrary
units) $D$ vs. $1/\protect\sqrt{\Upsilon }$, obtained directly from time
series such as in (a).}
\label{fig1}
\end{figure}

We first recall the basic experimental results, and Fig.~\ref{fig1}, which
is in part adapted from previously published work~\cite{danna-nature}, is
used to introduce the method and summarize the basic data. The dynamic
behavior of the perturbed granular medium is deduced from \textit{noise
measurements}. The granular noise, denoted $|\theta (f)|^{2},$ is obtained
by detecting the irregular motion of a torsion oscillator (Fig.~\ref{fig1}
a), deeply immersed into the perturbed granular material, and taking the
squared amplitude of the Fourier transform of the observed data series. The
granular medium is perturbed by shaking the container by well isolated taps,
or by continuous vibrations. An accelerometer on the container measures the
intensity of the perturbation, quantified by the reduced peak acceleration 
$\Gamma =a_{s}\omega _{s}^{2}/g,$ written in terms of the amplitude $a_{s}$
and the frequency $f_{s}=\omega _{s}/2\pi $ of a sinusoidal shaking, and the
acceleration of gravity $g.$ Below the ideal fluidization limit at $\Gamma
_{f}=1,$ and at low-frequency (below the natural frequency of the
oscillator), we observe $1/f^{2}$ noise spectra.

This low-frequency noise is the structural, or
\textit{configurational} noise. This is easily understood considering
tapping experiments, where after each tap the granular system
completely stops in a static configuration, which in turn determines
an angular position of the immersed oscillator. A series of such taps
drives the granular medium from one static (or jammed) configuration
to another, and the corresponding series of static (or
''low-frequency'') angular positions $\theta (t)$ of the immersed
oscillator probes this process, which displays $1/f^{2}$ noise. Before
going further we emphasize some experimental aspects: first, when
continuous vibrations are used, the low-frequency spectrum is the same
as the one seen in tapping experiments\cite{danna-nature}. This gives
us a very rapid and efficient method to measure the noise, and most of
the data are obtained in this way. Second, our measurements are
performed after the sample has been prepared in a reproducible
state. In particular, after having been poured into the container, the
granular medium is subject to very strong vibrations (large $\Gamma $)
whose intensity is progressively reduced until reaching the value of
$\Gamma $ where the measurements are performed. This discards the
strong compaction effects occurring during the first run on a loose
granular system, should we have initiated the measurement directly at
a low $\Gamma $ after pouring. The procedure in fact brings the
system directly into an almost stationary state which should
correspond to the reversible curve described in~\cite{exp-chicago}.

In order to have a deeper insight on the fluctuations of the rotation
angles, we have directly used the time series of $\theta (t)$ (one
example is shown in Fig.~\ref{fig1}a) to compute the structure factor
$S(\tau)=\langle (\theta (t+\tau )-\theta (t))^{2}\rangle $ for
different values of $\Gamma $. The obtained linear dependence with
$\tau $ allows us to determine a diffusion coefficient $D(\Gamma )$,
which features an $\exp (-B^{\prime }/\sqrt{\Gamma })$ dependence at
low $\Gamma $, as displayed in the inset of
Fig.~\ref{fig1}b. Moreover, the $1/f^{2}$ noise resulting from the
time series also gives important hints on the nature of the slow
glass-like granular dynamics: since $|\theta (f)|^{2}$ is proportional
to a diffusion coefficient, its inverse at a given frequency can be
considered to be proportional to an intrinsic configurational
relaxation time, $\tau_{x},$ i.e., $1/|\theta (f)|^{2}=C\tau
_{x}$. In order to obtain the constant of proportionality $C$, the
noise data are compared with susceptibility
measurements~\cite{danna-pre}, since a peak in the loss factor (i.e. a
peak in the tangent of the argument of the complex susceptibility)
arises when $\omega_{p}\tau_{x}=1$, where $f_{p}=\omega_{p}/2\pi
$ is the frequency at which the susceptibility is measured. This gives
approximately $C=2\pi 1500$ mrad$^{-2}$s.

The following important points can be deduced from the experimental
measurements. First, previous experiments~\cite{danna-prl} provide
evidence for a parameter of the form $\Upsilon =b\sqrt{\Gamma }/\omega
_{s}=b\sqrt{a_{s}/g}$, with $b$ a constant. This key \textit{control
parameter} determines the ''level'' of the low-frequency noise, as
shown in~\cite{danna-prl}. In other words, whatever amplitude $a_{s}$
and frequency $f_{s}$ of the perturbation is, provided $\Gamma <1$,
the noise only depends on $\Upsilon $. (In Fig.~\ref{fig1}b we set
$b=g^{1/2}$ and the control parameter is $\Upsilon
=\sqrt{a_{s}}$. Notice that, since we have not changed $g$,
experiments only prove that the control parameter is proportional to
the square root of the perturbation amplitude, i.e., $\Upsilon
\propto \sqrt{a_{s}}$). Another indication for a control parameter
like $\Upsilon $ comes from the direct measurement of the diffusion
coefficient (inset of Fig.~\ref{fig1}b) which shows an $\exp
(-B^{\prime }/\Upsilon )$ form, in agreement with the measurements of
$\tau _{x}$. The physical signification of this empirical control
parameter will be discussed in detail below.

Summarizing, from noise measurements as a function of $\Gamma $ at
different shaking frequencies $f_{s}$, and having determined the
relationship between $|\theta (f)|^{2}$ and $\tau _{x}$ from
susceptibility measurements, we obtain the curves shown in
Fig.~\ref{fig1}b, namely a plot of the configurational relaxation time
$\tau _{x}$ as a function of the control parameter $\Upsilon
$. Secondly, the data for small $\Upsilon $ in Fig.~\ref {fig1}b have
been fitted with the expression $\tau _{x}=A\exp [B/(\Upsilon
-\Upsilon _{0})^{p}],$ which gives in particular $p=0.95.$ Since the
observed exponent is close to $p=1,$ we assume from now on that the
configuration relaxation time is well described by the standard
Vogel-Fulcher-Tammann (VFT) expression
\begin{equation}
\tau _{x}\simeq A\exp [B/(\Upsilon -\Upsilon _{0})].  \label{VFT}
\end{equation}
In this VFT form, $\Upsilon $ is the empirical control parameter
playing the role of temperature (which does not mean that it
\textit{is} a temperature!), and $\Upsilon _{0}$ is the value of the
control parameter where the configuration relaxation time scale
diverges.\footnote{ Actually the data for $\tau _{x}$ can also be
fitted by the simple exponential form $\tau _{x}\simeq A\exp
[B/\Upsilon ]$, as well as by the form $A\exp \left[ B/\left(
\sqrt{\Upsilon ^{2}-\Upsilon _{0}^{2}}\right) \right] $, at least for
$\Upsilon \gg \Upsilon _{0};$ we shall come back to this point on the
discussion of the model.(An experimental estimate of $\Upsilon _{0}$
gives $1.37\times 10^{-4}$ $m^{1/2}$, which corresponds to a
displacement of about$\ 19$~$nm.$ Notice that $\Upsilon _{0}$ is very
sensitive to the exponent $p,$ and should be considered only as an
order of magnitude.)}

\section{Model}

Before going further, it is important to identify the granular regime
we are addressing. It is evident from Fig.~\ref{fig1}b that $\Upsilon
$ appears as the relevant control parameter only for external shaking
with $\Gamma <1.$ In fact, the scaling described by eq.(\ref{VFT})
ceases to hold for values of $\Upsilon $ which correspond, for each
$f_{s},$ to values of $\Gamma $ around unity. This is an important
point and the model is supposed to be valid for $\Gamma \leq 1$. In
this regime no fluidization occurs and the granular medium can be
considered in a \textit{quasi-solid} phase where geometric frustration
plays a major role. During a periodic vertical shaking of amplitude
$a_{s}$ and frequency $f_{s}$, with $\Gamma <1,$ the granular medium
moves as a whole, following the movements imposed to the container.
At the same time the torsion oscillator immersed in the medium is
fixed in the laboratory reference frame, i.e., it moves with respect
to the medium with the same periodicity imposed by the shaking. It
turns out that it is easier to explain what happens during the shaking
if one inverts the point of view and considers the granular medium as
globally immobile in the laboratory reference frame, while the torsion
oscillator is subject to a vertical oscillatory forcing. Thus, at each
period the torsion oscillator penetrates the granular medium for a
length $a_{s}$ and returns to the original position. In other words,
one externally imposes a given displacement (or deformation) to the
medium, while the stress exerted is not fixed.

What happens at the microscopic level? During the excursion of length
$a_{s}$ , the torsion oscillator will try to find its way inside the
granular medium by advancing and possibly rotating of a certain angle
in order to adapt to the reaction of the granular system. We have the
following picture. For small excursions $a_{s}$\ (to be defined below)
the medium stays in a given static (or jammed) configuration, and it
is deformed either elastically or plastically. Eventually the medium
\textit{fractures}, and jumps to another jammed configuration. Our
immersed oscillator detects essentially each fracture-event, and a
time-series represents a sequence of different jammed
configurations. In contrast to an ordinary solid, for which a fracture
is a unique fatal event, the granular solid is able to reestablish
itself in a new jammed configuration and to support successive
fractures.

Thus we assume that the penetration/deformation process involves three
distinct regimes: 1) An \textit{elastic} regime for excursions
$a_{s}$\ below a limit denoted $a_{s}^{0}$. For $a_{s}<a_{s}^{0}$\ the
system is able to absorb the imposed displacement elastically. (The
experimental value of $a_{s}^{0}$\ is material dependent and, as
discussed below, we identify it with the displacement for which we
observe the divergence at $\Upsilon _{0},$ \ i.e., $a_{s}^{0}\approx
10^{-8}$~m.) In this regime the granular system responds as an elastic
medium and the torsion oscillator returns to the original angular
position at the end of the cycle; 2) A \textit{plastic} regime, in
which the imposed displacement is absorbed by the granular system by
reorganizing the \emph{internal stress network}, inducing irreversible
rotations of the angular oscillator position, but without leaving the
actual jammed configuration; 3) A \textit{fracture} process in which
the system is unable to further absorb the imposed displacement, and a
\emph{macroscopic grain rearrangement} (or internal avalanche) is
required, resulting in a large jump of the torsion angle $\theta (t)$.

We now formalize this three regime process. In the plastic phase the
torsion angle undergoes sudden (but very small) changes due to the
reorganization of the internal stress distribution. In a first
approximation one can imagine that the variations of the torsion angle
could be described by a discrete random walk, i.e. one has a gaussian
distribution for the torsion angles whose variance has to be computed
by estimating the number of random walk steps the torsion oscillator
undergoes for an excursion of length $a_{s}$.  We shall see in the
following how the hypothesis of a gaussian distribution, partially
supported by the results for the diffusion coefficient, is not crucial
and can be relaxed without changing the main results.

Since the relaxation law (\ref{VFT}) implies that below a certain
value of $a_{s}$ the relaxation time is always infinite, we can deduce
the existence of an elastic threshold for $a_{s}$,
i.e. $a_{s}^{0}$. This means that for $a_{s}<a_{s}^{0}$ no permanent
deformation is produced in the system and the typical time for a
macroscopic rearrangement is infinite, hence the existence of an
\textit{elastic phase}. For $a_{s}>a_{s}^{0}$ the reaction of the
medium on the torsion oscillator induces irreversible rotations, hence
the existence of a \textit{plastic phase}. For these reasons
$a_{s}^{0} $ can be considered as the length of the elementary step of
the random walk.

Now, provided that $a_{s}>a_{s}^{0}$, the number of random walk steps
$n$ performed by the torsion oscillator for a displacement $a_{s}$
will be proportional to $a_{s}-a_{s}^{0}$. If the elementary angle of
rotation at each step is of the order of $\theta _{el}$ we expect a
gaussian distribution for the rotation angles with variance $\sigma
\propto \theta _{el}\sqrt{n}\propto \theta _{el}\sqrt{
(a_{s}-a_{s}^{0})/a_{s}^{0}}$. This is a first indication that a
parameter proportional to $\sqrt{a_{s}}\propto \Upsilon $ (or $\sqrt{
(a_{s}-a_{s}^{0})/a_{s}^{0}}\propto \sqrt{\Upsilon ^{2}-\Upsilon
_{0}^{2}}$ ) could play the role of a control parameter, i.e.  a role
similar to a temperature in the sense that it determines the variance
of the stress fluctuations induced indirectly by the advancing torsion
oscillator.

Let us now try to explain the VFT behavior of the relaxation
time. This time represents the typical time for the system to undergo
a macroscopic grain rearrangement. The question can be then rephrased
as follows: what is the probability that the torsion oscillator, in
its angular random walk with variance proportional to $\theta
_{el}\sqrt{ (a_{s}-a_{s}^{0})/a_{s}^{0}}$, will produce a macroscopic
configurational change such that the system jumps to another static
(jammed) configuration? Our hypothesis is that the fracture event
takes places as an \textit{extreme event} whose probability can be
computed in the framework of the extreme order statistics \cite
{galambos}. Fluctuations in the torsion angle correspond to a stress
redistribution and one expects larger angles to produce larger
compression of the grain-chains involved. It is natural to expect that
there will be some threshold value $\theta _{f}$ for the fluctuating
torsion angle above which the contact network will break and yield to
a macroscopic rearrangement leading to a large jump for $\theta
(t)$. We have then to look for the probability that, among the
fluctuations of the torsion angle excited at each shaking (i.e., at
each penetration over $a_{s}$), \textit{the largest fluctuation will
be larger than a given threshold} $\theta _{f}$.

We define $Z_{n}=\max (X_{1},X_{2},...,X_{n}),$ with $X_{i}=\theta
_{i}/\sigma $. We search the probability that $\Pr [Z_{n}>x_{f}]\equiv
S_{n}(x_{f})$, where $x_{f}=\theta _{f}/\sigma $ is the normalized
angular threshold. The (cumulative) probability distribution is $\Pr
[Z_{n}\leq x]\equiv H_{n}(x),$ and $S_{n}(x)=1-H_{n}(x).$ According to
standard text-books, for a Normal parent probability density
distribution, using the sequences given by $c_{n}=\sqrt{2\ln n}-(\ln
\ln n+\ln 4\pi )/(2\sqrt{2\ln n} )$ and $d_{n}=1/\sqrt{2\ln
n}$~\cite{galambos}, the probability distribution
$H_{n}(c_{n}+d_{n}x)$ tends, as $n$ increases, to a Gumbel probability
distribution~\cite{gumbel}, i.e. $\lim\limits_{n\rightarrow \infty
}H_{n}(c_{n}+d_{n}x)=\exp [-\exp (-x)]$. For large $n$ the probability
$S_{n}(x_{f})$ is given by $S_{n}(x_{f})=1-\exp \{-\exp
[-(x_{f}-c_{n})/d_{n}]\}$.

For $(x_{f}-c_{n})/d_{n}\gg 1,$ the probability $S_{n}(x_{f})$ can be
approximated by an exponential function, i.e., $S_{n}(x_{f})\approx
\exp [-(x_{f}-c_{n})/d_{n}]$. Using the expressions for $c_{n}$ and
$d_{n}$ given above, the above condition is verified if $n^{2}\exp
\left[ \frac{-\theta _{f}}{\sigma }\sqrt{2\ln n}\right] \ll \sqrt{4\pi
\ln n}$. In our case $\sigma \propto \theta
_{el}\sqrt{(a_{s}-a_{s}^{0})/a_{s}^{0}}$ and the approximation is
correct if $\theta _{f}/\theta _{el}\gg \sqrt{2n\ln n}-
\frac{\sqrt{n}\ln \sqrt{4\pi \ln n}}{\sqrt{2\ln n}}$, which is a
reasonable assumption given the experimental values of the
parameters. Using the previous approximation the probability for a
rare fracture event (corresponding to a configurational rearrangement)
is
\begin{equation}
S_{n}(\theta _{f}/\sigma )\approx \frac{n^{2}}{\sqrt{4\pi \ln n}}\exp \left[ 
\frac{-\theta _{f}}{\sigma }\sqrt{2\ln n}\right] .  \label{exp}
\end{equation}

\noindent We expect that $\theta _{f}$ will depend on the specific
tribological properties of the grains, as well as on some geometrical
properties of the system, such as the grain shape and size
distribution.  Thus, for a given granular system (i.e., fixed $n$ and
$\theta _{f}$) the fracture probability is determined only by $\sigma
$. The inverse of the probability $S_{n}(\theta _{f}/\sigma )$
determines the characteristic time for grain configuration
rearrangements, i.e. the characteristic time of the macroscopic
dynamics. One has then $S_{n}\propto \tau _{x}^{-1}$. Recalling that
$\sigma $ is directly related to the empirical control parameter
$\Upsilon \propto \sqrt{a_{s}}$, one finds
\begin{equation}
\tau _{x}\simeq A\exp \left[ B/\sqrt{\Upsilon ^{2}-\Upsilon _{0}^{2}}\right]
,
\end{equation}
which, given the extremely small experimental value of $a_{s}^{0}$, is
indistinguishable from~(\ref{VFT}) and also very close to an Arrhenius
behavior. As already stressed the experimental data do not allow to
definitely discriminate between the three behaviors.

It is important to notice that in the present model the gaussian
distribution of the torsion angle in the plastic regime follows from
the random walk analysis. However the precise form of the torsion
angle distribution is irrelevant as long as it decays faster than any
power-law (see~\cite{bouchaud}). For instance for stretched
exponential distributions as $exp{(- \alpha |\theta|^{\beta})}$ one
would obtain again a Gumbel distribution for the largest fluctuation
where the parameters $c_{n}$ and $d_{n}$ would be given, to
logarithmic accuracy, by $c_{n}\simeq {(2\ln n/\alpha)}^{1/\beta}$ and
$d_{n}=1/c_{n}$. This leads to the conclusion that, independently of
the precise form of the torsion angle distribution (i.e. under very
mild assumptions on it) one gets the same result for the relaxation
time provided the variance of the distribution is proportional to
$\sqrt{(a_{s}-a_{s}^{0})/a_{s}^{0}}$.

\section{Conclusions}

We have proposed a microscopic model which is able to explain some
experimental results for the relaxation dynamics of a granular medium
described~\cite{danna-nature,danna-prl}. The crucial hypothesis is
that at low $\Gamma$ ($\Gamma < \Gamma_f$ where $\Gamma_f \simeq 1$ in
our experiments) the macroscopic dynamics is controlled by extreme
events of the stress fluctuations in the system: if the variance of
these stress fluctuations, driven by the imposed penetration of the
oscillator over the distance $a_{s}$, is proportional to $\sqrt{
(a_{s}-a_{s}^{0})/a_{s}^{0}}$ (as it turns out if the microscopic
torsion angle fluctuations are Gaussian), then a macroscopic change of
$\theta (t)$ can occur only if such microscopic fluctuations overcome
a threshold, i.e. if an extreme fracture event takes place. These
simple ingredients immediately lead to an activated-like, VFT-like
character of the relaxation time. The deviation of the relaxation
behaviour from the Arrhenius law is explained by the existence of an
\textit{elastic threshold} $a_{s}^{0}$, which is material dependent:
for $a_{s}<a_{s}^{0}$ the system can absorb the perturbation without
any rearrangement.

For $\Gamma >\Gamma _{f}$ the situation is quite different: at each
cycle most of the induced fluctuations in the system are large enough
for a macroscopic rearrangement to occur, and the dynamics is not
controlled anymore by rare events. In this fluid-like phase,
structural (configurational) changes are not rare events. Recently
Philippe and Bideau have identified, using a geometry very similar to
the one of our experiments, a threshold value of the order of $1.2$
above which the system is able to reach a stationary state with a
density relaxation ruled by a stretched exponential
law~\cite{rennes}. From this point of view we speculate that the
threshold value for fluidization could also mark the boundary between
a glassy region (low $\Gamma $) where the relaxation is driven by
extreme isolated events and a quasi-liquid region (high $\Gamma $)
where the system is able to reach a stationary state. It is important
to stress that the threshold value for fluidization can depend on the
geometry of the container. In particular we expect the threshold value
to increase a lot for highly confined geometries where the effect of
the boundaries is strong. This could be the case for the Chicago
experiments~\cite{exp-chicago} where a narrow and tall container is
used. If the effective threshold for this experiment was large (for
instance $\Gamma_f \simeq 3\div 4$) this could explain why they
observed a very slow relaxation even for values of $\Gamma$ well above
$1$.

The present work represents only a first step in the direction of a better
link between the microscopic dynamics of granular media and the macroscopic
response to an external perturbation. From this point of view, various
experiments can be thought of to clarify several points: the actual
functional form for the relaxation time as well as the dependence of the
elastic threshold $a_{s}^{0}$ on the material, the dependence of the control
parameter on the acceleration of gravity $g$, the exploration of the
fluid-like regime ($\Gamma > \Gamma_f$). Finally the experimental setup
described in this paper could allow the measurement of some thermodynamical
properties of granular media: in particular the existence of effective
temperatures could be investigated by a suitable combination of
susceptibility and noise measurements, in the spirit of recent experiments
on laponite~\cite{Bellon}. Such experiments would open a way towards a
comparison with recent theoretical predictions on a thermodynamical approach
to granular matter~\cite{Edwards,bklm}.

\acknowledgements This work is supported by the Swiss National Science
Foundation as well as and INFM \emph{Center for Statistical Mechanics and
Complexity} (SMC).


\begin{thebibliography}{99}
\bibitem{RMP}  H.M. Jaeger, S.R. Nagel, and R.P. Behringer, \emph{Rev. Mod.
Phys} {68}, 1259 (1996).

\bibitem{exp}  \emph{Physics of Dry Granular Media} (eds. Herrmann, H. J.,
Hovi, J.-P. and Luding, S.) 553-583 (Kluwer Academic, Dordrecht, The
Netherlands, 1998).

\bibitem{exp-chicago}  Knight, J. B., Fandrich, C. G., Lau, C. N., Jaeger,
H. M. and Nagel, S. R. Phys. Rev. E \textbf{51}, 3957 (1995); Nowak, E. R.,
Knight, J. B., Ben-Naim, E., Jaeger, H. M. and Nagel, S. R., Phys. Rev. E 
\textbf{57}, 1971-1982 (1998).

\bibitem{jamming}  See the collection of papers in \emph{Jamming and
Rheology: Constrained Dynamics on Microscopic and Macroscopic Scales}, ed.
by A.J. Liu and S.R. Nagel (Taylor \& Francis, New York 2001).

\bibitem{danna-nature}  G. D'Anna and G. Gremaud, \emph{Nature}, \textbf{413}
, 407 (2001).

\bibitem{danna-prl}  G. D'Anna and G. Gremaud, \emph{Phys. Rev. Lett.}, 
\textbf{87}, 254302 (2001).

\bibitem{danna-pre}  G. D'Anna and G. Gremaud, \emph{Phys. Rev. E}, 
\textbf{64}, 011306 (2001).

\bibitem{rammal}  R. Rammal, \emph{J. Physique}, \textbf{46}, 1837 (1985).

\bibitem{vinokur}  V. Vinokur, M. Cristina Marchetti and L.W. Chen, \emph{\
Phys. Rev. Lett.} \textbf{77}, 1845 (1996).

\bibitem{bouchaud}  J.-P. Bouchaud and M. M\'{e}zard, \emph{J. Phys. A} 
\textbf{30}, 7997 (1997).

\bibitem{galambos}  Galambos J., \emph{The Asymptotic Theory of Extreme
Order Statistics}, 2nd ed., Krieger, (1987).

\bibitem{gumbel}  Gumbel E.J., \emph{Statistics of Extreme}, Columbia
University Press, New York, NY. (1958).

\bibitem{rennes}  P. Philippe and D. Bideau, to appear in
\emph{Europhys. Lett.} (2002).

\bibitem{Bellon} L. Bellon and S. Ciliberto, \emph{Europhys. Lett.}
\textbf{53}, 511 (2001); \emph{Physica D} \textbf{168-169}, 325
(2002).

\bibitem{Edwards}  Edwards, S.F. \emph{The Role of Entropy in the
Specification of a Powder}, in:\textit{Granular Matter: An Interdisciplinary
Approach}, A. Mehta, Ed. (Springer-Verlag, New York, 1994), and references
therein. S. F. Edwards, \emph{J. Stat. Phys.} \textbf{62} (1991) 889; Mehta
A. and Edwards S. F., \emph{Physica A} \textbf{157} (1989) 1091.

\bibitem{bklm}  A. Barrat, J. Kurchan, V. Loreto, M. Sellitto, \emph{Phys.
Rev. Lett.} \textbf{85}, 5034 (2000); \emph{Phys. Rev. E} \textbf{63},
051301 (2001).

\end{thebibliography}
\end{document}